# Electronic Structure and Small Hole Polarons in YTiO$_3$

Jin Yue[1,a)], Nicholas F. Quackenbush[2], Iflah Laraib[3], Henry Carfagno[3], Sajna Hameed[4], Abhinav Prakash[1], Laxman R. Thoutam[1], James M. Ablett,[5] Tien-Lin Lee,[6] Martin Greven[4], Matthew F. Doty[3], Anderson Janotti[3], and Bharat Jalan[1,a)]

[1]Department of Chemical Engineering and Materials Science

University of Minnesota,

Minneapolis, Minnesota 55455, USA

[2]Materials Measurement Science Division, Material Measurement Laboratory

National Institute of Standards and Technology,

Gaithersburg, MD 20899, USA

[3]Department of Materials Science and Engineering

University of Delaware,

Newark, Delaware 19716, USA

[4]School of Physics and Astronomy

University of Minnesota,

Minneapolis, Minnesota 55455, USA

[5]Synchrotron SOLEIL, L'Orme des Merisiers, Boîte Postale 48, St Aubin, 91192 Gif sur Yvette, France

[6]Diamond Light Source, Ltd., Harwell Science and Innovation Campus,

Didcot, Oxfordshire OX11 0DE, England, UK

a) Corresponding author: yuexx129@umn.edu, bjalan@umn.edu

**Abstract**

As a prototypical Mott insulator with ferromagnetic ordering, YTiO$_3$ (YTO) is of great interest in the study of strong electron correlation effects and orbital ordering. Here we report the first molecular beam epitaxy (MBE) growth of YTO films, combined with theoretical and experimental characterization of the electronic structure and charge transport properties. The obstacles of YTO MBE growth are discussed and potential routes to overcome them are proposed. DC transport and Seebeck measurements on thin films and bulk single crystals identify *p*-type Arrhenius transport behavior, with an activation energy of ~ 0.17 eV in thin films, consistent with the energy barrier for small hole polaron migration from hybrid density functional theory (DFT) calculations. Hard X-ray photoelectron spectroscopy measurements (HAXPES) show the lower Hubbard band (LHB) at 1.1 eV below the Fermi level, whereas a Mott-Hubbard band gap of ~1.5 eV is determined from photoluminescence (PL) measurements. These findings provide critical insight into the electronic band structure of YTO and related materials.

The coupling among charge, orbital, lattice, and spin degrees of freedom in perovskites $ABO_3$ with partially filled $d$-bands can lead to a wide range of fascinating properties like colossal magnetoresistance and exotic magnetic behavior.[1] Among the $ABO_3$ perovskites, the rare earth titanates $R$TiO$_3$ ($R$ = rare earth elements, except Eu) are particularly interesting, as they are prototypical Mott insulators and provide a perfect playground for the study of strong electron-correlation effects.[2] One of many peculiar features of this isostructural series of compounds is the magnetic ground-state transition from ferromagnetism or ferrimagnetism in $R$TiO$_3$ with smaller $R$ ions (e.g., YTiO$_3$, or YTO) to antiferromagnetism in those with larger $R$ ions (e.g., LaTiO$_3$).[2-4] For most $R$TiO$_3$, the partially-filled 4$f$-shell of the rare-earth ions leads to a large total spin angular momentum, and hence adds complexity to the magnetic properties, as the spin degrees of freedom of both Ti$^{3+}$ and rare-earth ions can lead to complex magnetic ordering. However, this complexity can be circumvented in YTO, since Y$^{3+}$ does not contain $f$ electrons.

Unlike most Mott insulators, YTO shows a ferromagnetic ground state, below ~ 30 K.[5,6] The origin of this ferromagnetic state has been the subject of numerous theoretical studies[7-10] and experimental investigations, including inelastic X-ray scattering[11,12], polarized neutron diffraction[13-15], photoemission spectroscopy[16-18], and Raman scattering[19]. Despite these extensive research efforts, many important questions remain unanswered. For instance, what is the governing transport mechanism in YTO at room temperature? Similarly, the value of the Mott-Hubbard bandgap, $E_g$ in YTO is not settled. Early attempts have focused on optical conductivity ($\sigma(\omega)$) measurements to determine $E_g$.[20-24] The onset in $\sigma(\omega)$ spectrum was attributed to interband transitions. Typical $\sigma(\omega)$ spectra showed two onsets, one at < ~ 1 eV and another at ~ 4 eV, which were attributed to Mott-Hubbard gap transitions from the lower Hubbard band (LHB) to the upper Hubbard band (UHB) ($|d^1d^1\rangle \rightarrow |d^0d^2\rangle$) and charge-transfer gap transitions from occupied O 2$p$ states to Ti 3$d$ UHB ($|d^1p^6\rangle \rightarrow |d^2p^5\rangle$), respectively.[20-24] More recently, Gössling *et al.* examined ellipsometry data and argued that the lowest electronic transition across $E_g$ occurs at 2.55 eV.[25] Ellipsometry data by Kovaleva *et al.* further support this argument, giving a slightly higher transition energy of 2.85 eV.[26] In contrast, recent DFT calculations with intra-site Coulomb repulsion $\underline{U}$ predicted $E_g \sim 2$ eV and attributed the low-energy onset in the optical conductivity to excitations of electrons from the LHB to empty (hole) polaronic states.[8]

In order to gain deeper insight into the electronic structure of YTO, we synthesized films using hybrid molecular beam epitaxy (h-MBE) method and investigated the electronic properties of YTO through a combination of hybrid density functional calculations and experimental characterization techniques. We also grow single cyrstals with the floating-zone technique for some complementary measurements. DC transport and Seebeck measurements reveal that YTO shows *p*-type Arrhenius transport behavior, with an activation energy of 0.17 eV for the film samples. Based on hybrid functional calculations, we attribute this activation energy to a small hole polaron migration barrier. Our hard X-ray photoemission spectroscopy (HAXPES) measurements revealed a LHB at 1.1 eV below the Fermi level, and photoluminescence (PL) measurements indicate a Mott-Hubbard gap of ~1.5 eV, which is significantly larger than the onset at ~1 eV often observed in optical conductivity spectra. In the following, we describe the experimental and theoretical approaches, paying special attention to the obstacles in the MBE growth of YTO, and then report the main results and conclusions of this combined theory-experiment effort.

First, we emphasize that MBE growth of YTO films has not been reported before, to the best our knowledge. The main challenge is associated with the evaporation of Y. Since the discovery of the high-temperature superconductor $YBa_2Cu_3O_{7-x}$ (YBCO), researchers have been trying to grow yttrium-containing compounds with MBE. However, due to the extremely low vapor pressure of Y ($2\times10^{-5}$ mbar at 1540°C)[27], the majority of the research efforts focused on e-beam evaporation, which has its own challenges. One of the major challenges of e-beam evaporation is maintaining a stable beam flux, owing to local heating behavior.[28] In the present study, we employed a thermal effusion cell with a tantalum crucible to thermally evaporate/sublimate Y. Ta was chosen owing to its oxygen resistance, high malleability and prior use as a crucible material in e-beam sources.[29]

Our initial growth attempts failed - after about 40 thermal cycles, the crucible cracked. We first discuss the lesson learned from this failure. Figure 1a shows an image of a cracked Ta-crucible for which the bottom part, where Y was in contact with the crucible, was found shrunk compared to the top part. We can understand this behavior by considering the large linear thermal expansion mismatch between Y ($\alpha_L = 10.6 \times 10^{-6}$/K) and Ta ($\alpha_L = 6.3 \times 10^{-6}$/K), which results in thermal stresses during heating/cooling cycles and, eventually, failure. At first, this result was quite surprising to, given that the effusion cell was operated (between 1400 – 1480 °C) below the melting point of pure Y (~ 1522 °C), and therefore, Y melting (and hence, the thermal stresses) was not expected. Looking at the Y-Ta phase diagram, we however found a nonzero solid-solubility of Ta in Y and a eutectic point at ~99.6 at.% Y + ~0.4 at.% Ta, which may explain the melting of Y at lower temperatures.[30] An apparent solution of this challenging problem would thus be to operate the effusion cell at lower temperatures, which is unfortunately not possible due to the low vapor pressure of Y. Our proposed solution to this issue was therefore to avoid thermal cycling by keeping the idle temperature of the cell just below the operating temperature (where vapor pressure is low). The effusion cell can be cooled down slowly to room temperature only after the source material is completely depleted. It may also be possible to avoid thermal stress by using a liner (such as $Al_2O_3$). Using the former approach, YTO films were successfully grown in the hybrid MBE set-up. [31-35]

Phase-pure, epitaxial, 12 nm thick YTO films were grown on $DyScO_3$ (DSO) (110) substrates using hybrid molecular beam epitaxy (MBE) approach (details described elsewhere[31-35]). All films were grown at a fixed substrate temperature of 950 °C (thermocouple temperature). No additional oxygen was used during growth because our source of Ti also supplies oxygen.[31-35] The growth was monitored *in-situ* using Reflection High-Energy Electron Diffraction (RHEED). High-resolution wide-angle X-ray diffraction (WAXRD) scans were used to characterize the samples' structural quality *ex-situ*. Bulk YTO crystals were grown using the floating-zone method.[20] Electrical characterization was performed in a Quantum Desing, Inc. Physical Property Measurement System (PPMS Dynacool) with Van der Pauw geometry. 20nm Ti/100 nm Au was sputtered on the corners of the samples as Ohmic contacts, and indium was soldered on top of the Au to make connections to the PPMS resistivity puck. Seebeck measurements were carried out to characterize charge-carrier type. For Seebeck measurements, a bulk YTO sample was mounted using silver paste and Cu leads in a four-terminal configuration, where two leads at the ends of a bar-shaped sample were used to pass current, and the middle two leads were used to measure the temperature difference $\Delta T$ and Seebeck voltage $\Delta V$ simultaneously. A square wave heat pulse was applied to the sample, and the measured $\Delta T$ and $\Delta V$ were modeled to extract the Seebeck coefficient using a thermal transport module from Quantum Design.

The valence band (VB) spectra of YTO films were characterized *via* HAXPES, with an excitation energy of 4 keV and using the high-resolution Si (220) channel-cut post monochromator. Slits were closed down to a 0.1 mm gap vertically to minimize beam flux at the sample and reduce sample charging. The analyzer was located at 90° from the incident beam and the sample is at 5° off grazing, giving a take-off-angle of 85°. The analyzer was set to 200 eV pass energy. This gives an overall resolution of 250 meV, as determined by the width of the Fermi edge of Au foil. The binding-energy axis was calibrated using both the Fermi edge and Au 4*f* core lines from Au foil. The photoluminescence (PL) measurements were carried out using a 532 nm excitation laser illuminating the sample at a glancing angle to ensure excitation of a region much larger than the PL collection volume. The emitted PL was collected by a lens and analyzed with a grating monochrometer and liquid-nitrogen-cooled Si CCD with a resolution limit of approximately 70 μev.

Our first-principles calculations were based on DFT[36,37] and the screened hybrid functional of Heyd, Scuseria, and Enzerhof (HSE06)[38,39] as implemented in the Vienna Ab initio Simulation (VASP) code[40,41]. The projector augmented wave (PAW) method was used to describe the interaction between the core and valence electrons[42]. Equilibrium lattice parameters of YTO were obtained by optimizing an orthorhombic primitive cell consisting of 20 atoms, with a 6×6×4 Γ-centered *k*-point mesh for integrations over the Brillouin zone, and a 500 eV cutoff for the planewave basis set. For the calculations of the small hole polaron in YTO, we used a supercell consisting of 160 atoms (2×2×2 repetition of the unit cell), and a single *k*-point, (1/4,1/4/1/4), for integrations over the Brillouin zone. All the atomic forces were minimized until they reached a maximum threshold of 0.01 eV/Å.

First, we present results from the structural characterization of our films. Figure 1 shows the high-resolution 2$\theta$-$\omega$ scan of a representative 12 nm YTO/DSO (110) sample, consistent with expectations for an epitaxial and phase-pure film. The film peak is relatively broad due to the small film thickness. The analysis of the film peak yielded an out-of-plane lattice parameter of 3.900 ± 0.002 Å, which is close to the (110)$_o$ interplanar spacing of the bulk YTO crystal, 3.881 Å. This observation is consistent with a nominally stoichiometric $(110)_o$-oriented YTO films on $(110)_o$ DSO substrate where subscript "o" refers to orthorhombic structure of YTO. The streaky RHEED pattern of the film after growth is shown in the inset of figure 1b, indicative of a smooth surface morphology.

We now turn to the discussion of charge transport results. The temperature-dependent resistivity of the YTO film shown in figure 2a indicates insulating transport behavior. Data for a YTO bulk sample are shown for comparison in the inset of figure 2a. An Arrhenius fit to $\rho(T) \sim e^{-E_a/kT}$ yields activation energies of $E_a \sim 0.17$ eV for and $\sim 0.31$ eV for the film and bulk sample, respectively. These values are slightly different from the previously reported bulk value of $\sim 0.25$ eV [3,21,43], probably due to slight differences in the amount of non-intentional defects in bulk and/or residual strain/non-stoichiometry in films. Seebeck measurements were performed on the bulk YTO sample to investigate the carrier-type. Figure 2b shows the same sign for $\Delta T$ and $\Delta V$, indicating *p*-type carriers, in agreement with the previous reports. [3,43] Note that the activation energy observed in YTO is significantly smaller than the predicted Mott-Hubbard gap and the observed features in the optical conductivity. In order to understand the origin of this small activation energy, we show in figure 2c the calculated atomic configuration and charge-density isosurface of a small hole polaron in YTO at a single Ti site (site A or B), as well as the hopping steps between adjacent Ti sites (site A to B). The calculated polaron migration barrier energy was 0.27 eV, in reasonably good agreement with the experimentally-determined transport activation energy $E_a$, indicating that the thermally-activated behavior in YTO is likely governed by small hole polaron hopping.

Finally, to obtain further insight into the electronic structure, figure 3a shows the VB spectrum of the same 12 nm YTO/DSO (110) sample using the HAXPES. The data close to Fermi level ($E_F$) was magnified 50 times to clearly show the LHB. These results show that the LHB is around 1.1 eV below $E_F$, consistent with previous resonant photoemission spectroscopy data by Arita et al. on bulk YTO samples.[16] We also estimate the bandwidth of the LHB ($W$) to be ~ 1.8 eV. The calculated density of states (DOS) for YTO is also shown in figure 3a. The HAXPES result matches reasonably well with the hybrid functional calculation, especially with respect to the separation between the O 2$p$ bands and the LHB composed of Ti 3$d$ orbitals. The calculated Mott-Hubbard gap of YTO is 2.0 eV, in agreement with previous calculations[8]. DFT calculations of the DSO substrate suggest the broad peak at higher binding energies, ~ - 11 eV is likely related to the underlying substrate. Other possibilities related to surface impurities/contamination may also be responsible for a measurable DOS in this energy range. Unfortunately, the DSO reference sample was too insulating when trying to measure the VB. Finally, in an attempt to determine this gap experimentally, we carried out PL measurements. The raw data of PL measurements for a LAO substrate and a 10 nm YTO sample on LAO substrate are shown in figure 3b. It is noted that films grown on LAO substrates were used for PL measurements because the DSO substrate yielded a strong background signal in PL, making the interpretation of the data difficult. Figure 3b shows a clear PL signal from the YTO film around 1.5 eV. The PL from the pure YTO film obtained by subtracting the substrate signal, is shown in the inset of figure 3b. We find a gap of ~1.5 eV, which is slightly smaller than the Mott-Hubbard gap from DFT.

In summary, we have demonstrated the first MBE growth of YTO films. Several challenges associated with the evaporation of Y through effusion cells and plausible ways to overcome these challenges are discussed. The electronic transport and Seebeck measurements of both films and bulk samples revealed a $p$-type, with a thermally-activated transport behavior. Combined with hybrid functional calculations, we found that the transport in YTO is governed by small-polaron hopping, with the calculated polaron migration energy barrier of 0.27 eV consistent with the experimental results. From HAXPES measurements, the LHB of YTO was found to be located 1.1 eV below the Fermi level. Using PL measurements, we determined a Mott-Hubbard bandgap of ~1.5 eV in the YTO films. This study addresses key MBE growth challenges, and we expect it to facilitate future investigations of electronic and magnetic ground states in strain-engineered heterostructures of MBE-grown YTO films.

**Acknowledgements:**

This work was primarily supported by the U.S. Department of Energy through the University of Minnesota Center for Quantum Materials, under Award No. DE-SC0016371. Parts of this work were carried out at the Minnesota Nano Center, which is supported by the National Science Foundation through the National Nano Coordinated Infrastructure (NNCI) under Award Number ECCS-1542202. Structural characterizations were carried out at the University of Minnesota Characterization Facility, which receives partial support from NSF through the MRSEC program. IF and AJ acknowledge support from the NSF Early Career Award Grant No. DMR-1652994. HC and MD acknowledge support from NSF DMR-1839056. It made use of the computing resources provided by the Extreme Science and Engineering Discovery Environment (XSEDE), supported by the National Science Foundation grant number ACI-1053575. The HAXPES measurements were performed while N.F.Q. held a National Institute of Standards and Technology (NIST) National Research Council (NRC) Research Postdoctoral Associateship Award at the Material Measurement Lab. We thank Diamond Light Source for access to beamline I-09 (SI17449-1) that contributed to the results presented here.

**Figures (Color Online):**

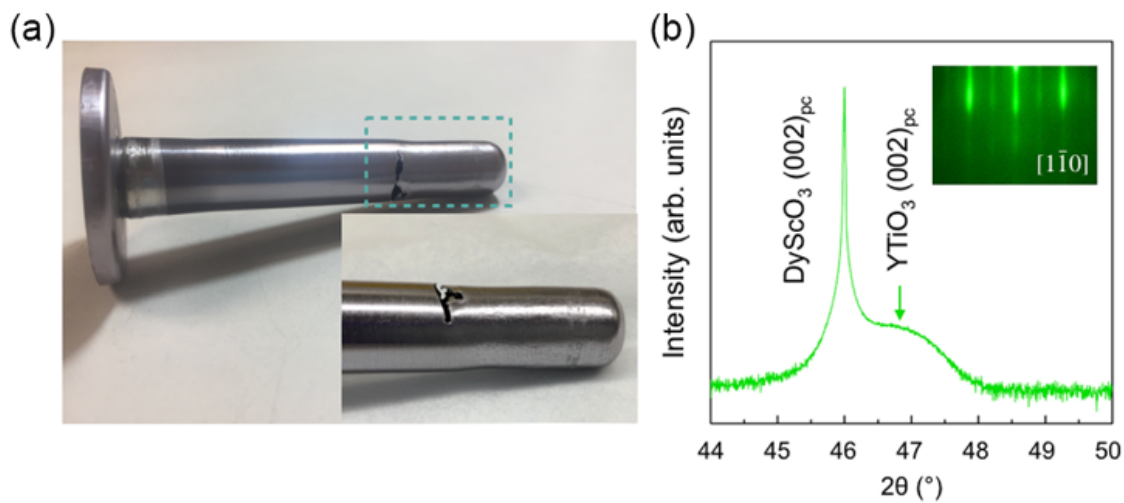

**Figure 1:** (a) Image of a cracked Ta crucible. Inset shows a zoomed-in picture around the crack (indicated by the green rectangle), where a clear shrinkage in volume can be observed at the bottom part of the crucible. (b) High-resolution x-ray $2\theta$-$\omega$ couple scan of a 12 nm YTO/ DSO (110) sample. Inset shows a RHEED pattern of the film after growth along the $[1\bar{1}0]$ azimuth of the substrate.

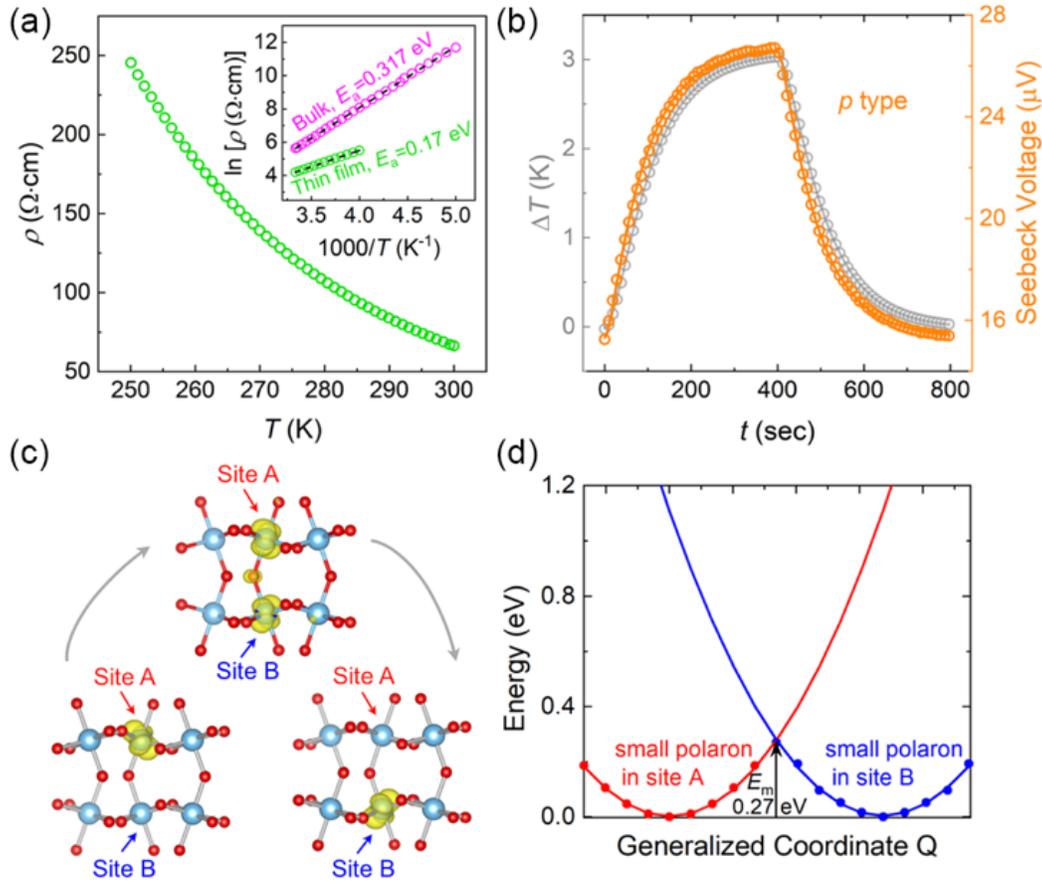

**Figure 2**: (a) $\rho$ vs. $T$ plot for a 12 nm YTO/DSO (110) sample. Inset: Arrhenius plot for this film and for a bulk YTO single crystal, along with the corresponding activation energies. (b) Seebeck measurement of the bulk YTO single crystal. A square-wave heat pulse was applied, then the temperature and voltage difference between the hot and cold leads were measured. Seebeck voltage (orange) yields the same sign as the temperature differences (grey), confirming $p$-type conduction in YTO. (c) Hopping process schematic of how a small polaron at Ti site $A$ migrates to neighboring Ti site $B$ via an intermediate state. (d) Configuration diagram for the polaron at adjacent Ti sites A and B. The calculated barrier energy for polaron migration is 0.27 eV.

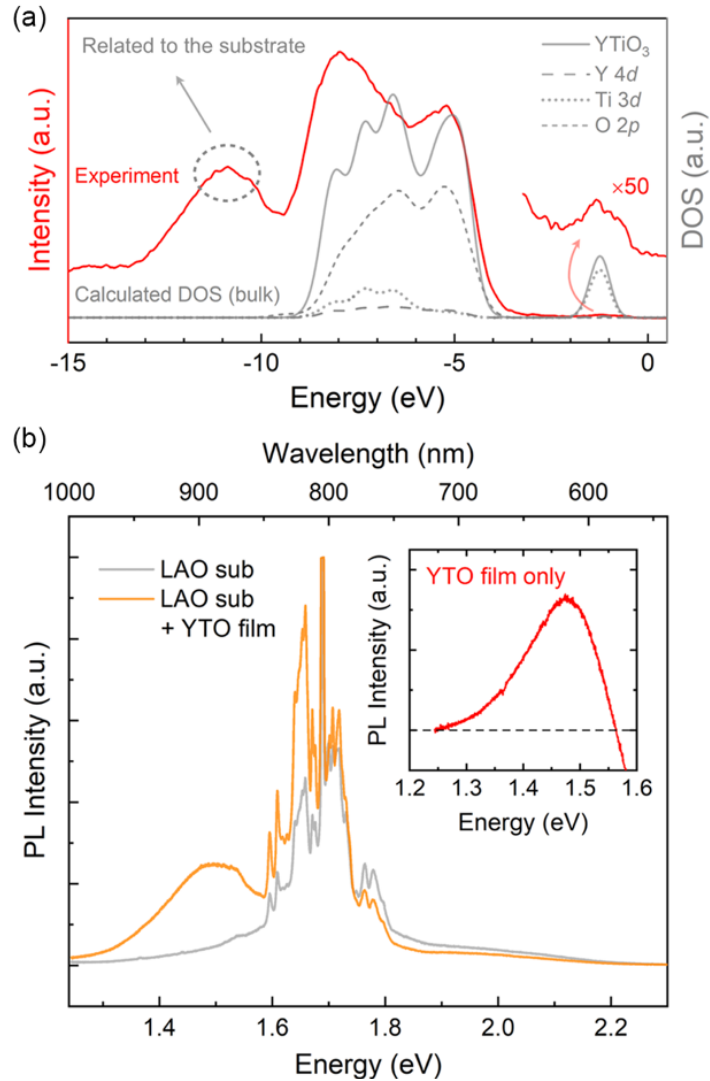

**Figure 3**: (a) HAXPES data (red color) from a 12 nm YTO/DSO (110)) along with the calculated density of states (grey color) for bulk YTO. The HAXPES data close to the Fermi level are shown on ×50 scale as an inset for clarity. (b) PL measurement of a YTO/LAO sample and a LAO substrate, where a clear signal from YTO film is observed around 1.2-1.6 eV. The inset shows the PL signal from the YTO film, which is obtained by subtracting the PL signal of the substrate from that of the substrate plus film.

15